\documentclass[letterpaper,11pt]{article}

\usepackage{verbatim}
\usepackage{latexsym}
\usepackage{amsmath,amssymb,amsthm}
\usepackage{fullpage}

\let\myPushQED=\pushQED
\let\myPopQED=\popQED
\newcommand{\myignore}[1]{}
\newenvironment{proof*}
  {\let\pushQED=\myignore\begin{proof}\let\pushQED=\myPushQED}
  {\def\popQED{}\end{proof}\let\popQED=\myPopQED}

\newenvironment{description*}%
  {\vspace{-1ex}\begin{description}%
    \setlength{\itemsep}{-0.5ex}%
    \setlength{\parsep}{0pt}}%
  {\end{description}}
\newenvironment{itemize*}%
  {\vspace{-1ex}\begin{itemize}%
    \setlength{\itemsep}{-0.5ex}%
    \setlength{\parsep}{0pt}}%
  {\end{itemize}}
\newenvironment{enumerate*}%
  {\vspace{-1ex}\begin{enumerate}%
    \setlength{\itemsep}{-0.5ex}%
    \setlength{\parsep}{0pt}}%
  {\end{enumerate}}

{\makeatletter
 \gdef\xxxmark{%
   \expandafter\ifx\csname @mpargs\endcsname\relax 
     \expandafter\ifx\csname @captype\endcsname\relax 
       \marginpar{xxx}
     \else
       xxx 
     \fi
   \else
     xxx 
   \fi}
 \gdef\xxx{\@ifnextchar[\xxx@lab\xxx@nolab}
 \long\gdef\xxx@lab[#1]#2{{\bf [\xxxmark #2 ---{\sc #1}]}}
 \long\gdef\xxx@nolab#1{{\bf [\xxxmark #1]}}
}

\newtheorem{theorem}{Theorem}
\newtheorem{lemma}[theorem]{Lemma}

\newcommand{\poly}{\mathrm{poly}}

\newcommand{\eps}{\varepsilon}
\newcommand{\twodots}{\mathinner{\ldotp\ldotp}}
\newcommand{\proc}[1]{\textnormal{\scshape#1}}

\newcommand{\rank}{\proc{Rank}}
\newcommand{\ans}{\mathrm{Ans}}
\newcommand{\probes}{\mathrm{Probes}}
\newcommand{\foot}{\mathrm{Foot}}

\newcommand{\pub}{\mathcal{P}}
\let\phi=\varphi

\newcommand{\E}{\mathbf{E}}
\newcommand{\evt}{\mathcal{E}}
\newcommand{\fixQ}{\widetilde{Q}}

\title{A Lower Bound for Succinct Rank Queries}
\author{
    Mihai P\v{a}tra\c{s}cu \\ IBM Almaden
}

\begin{document}

\maketitle

\begin{abstract}
  The rank problem in succinct data structures asks to preprocess an
  array $A[1\twodots n]$ of bits into a data structure using as close
  to $n$ bits as possible, and answer queries of the form $\rank(k) =
  \sum_{i=1}^k A[i]$. The problem has been intensely studied, and
  features as a subroutine in a majority of succinct data structures.

  We show that in the cell probe model with $w$-bit cells, if rank takes
  $t$ time, the space of the data structure must be at least $n +
  n/w^{O(t)}$ bits. This redundancy/query trade-off is essentially
  optimal, matching our upper bound from [FOCS'08].
\end{abstract}


\section{Introduction}

\subsection{The Complexity of Rank}

Consider an array $A[1\twodots n]$ of bits. Can we preprocess this
array into a data structure of size $n + r$ bits, for small redundancy
$r$, which supports rank queries $\rank(k) = \sum_{i=1}^k A[i]$
efficiently? The problem of supporting rank (and the related select
queries) is the bread-and-butter of succinct data structures. It finds
use in most other data structures (for representing trees, graphs,
suffix trees / suffix arrays etc), and its redundancy / query
trade-off has come under quite a bit of attention.

Rank already had a central position in the seminal papers on succinct
data structures. Jacobson \cite{jacobson89succinct}, in FOCS'89, and
Clark and Munro \cite{clark96succinct}, in SODA'96, gave the first
data structures using space $n+o(n)$ and constant query time. These
results were slightly improved in \cite{munro96succinct,
  munro98succinct, raman02perfhash}.

In several applications, the set of ones is not dense in the
array. Thus, the problem was generalized to storing an array
$A[1\twodots u]$, containing $n$ ones and $u-n$ zeros. The optimal
space is $B = \lg \binom{u}{n}$. Pagh~\cite{pagh01redundancy} achieved
space $B + O(n\cdot \frac{(\lg\lg n)^2}{\lg n})$ for this sparse
problem. Recently, Golynski et al.~\cite{golynski07rank} achieved $B +
O(n\cdot \frac{\lg\lg u}{\lg^2 n})$. Subsequently, Golynski et
al.~\cite{golynski07rank-b} have achieved space $B + O(n\cdot
\frac{\lg\lg n \cdot \lg(u/n)}{\lg^2 n})$.

In my paper from FOCS'08 \cite{patrascu08succinct}, I gave a
qualitative improvement to these bounds, showing an exponential
dependence between the query time and the redundancy. Specifically,
with query time $O(t)$, the achievable redundancy is $r \le n /
(\frac{\lg n}{t})^t$. This improved the redundancy for many succinct
data structures where rank/select queries were the bottleneck.

Given the surprising nature of this improvement, a natural question is
whether we can do much better. In this paper, we show that we cannot,
at least for the basic rank queries:

\begin{theorem}
In the cell-probe model with words of $w \ge \lg n$ bits, a data
structure that supports rank queries in $t$ cell probes requires 
at least$n + n / w^{O(t)}$ bits of space.
\end{theorem}

All succinct data structure papers assume $w = \lg n$. The lower bound
matches my upper bound, except for the difference between $(\lg n)^t$
and $(\frac{\lg n}{t})^t$. This difference is inconsequential for
small $t < \lg^{0.99} n$. If we want a polynomially small redundancy
(say, less than $n^\alpha$, for some constant $\alpha < 1$), the upper
bound says that $t = O(\lg n)$ is sufficient. The lower bound says
that $t = \Omega(\lg n / \lg\lg n)$ is necessary. It is unclear which
bound is the optimal one in this regime.

\subsection{Lower Bounds for Succinct Data Structures}

Much work in lower bounds for succinct data structures has been in the
so-called systematic model. In this model, the array $A$ must be
represented as is, i.e.~the data structure only has oracle access to
it (it can read any $w$ consecutive bits at $O(1)$ cost). In addition,
the data structure may store an index of sublinear size, which the
query algorithm can examine at no cost. See \cite{gal03succinct,
  miltersen05succinct, golynski07rank-b, golynski07succinct} for
increasingly tight lower bounds in this model. Note, however, that in
the systematic model, the best achievable redundancy with query time
$t$ is $\frac{n}{t\cdot \poly\lg n}$, i.e.~there is a linear trade-off
between redundancy and query time. This is significantly improved by
my (non-systematic) upper bounds \cite{patrascu08succinct}, and these
lower bounds qualitatively miss the nature of this improvement.

In the unrestricted cell-probe model, the first lower bounds were
shown by G\'al and Miltersen~\cite{gal03succinct} in 2003. These lower
bounds were strong, showing a linear dependence between the query and
redundancy $r \cdot t = \Omega(n/\lg n)$. However, the problem being
analyzed is somewhat unnatural: the bound applies to polynomial
evaluation, for which nontrivial succinct upper bounds appear
unlikely. Their technique, which is based on the strong error
correction implicit in their problem, remains powerless for ``easier''
problems. (Thus, succinct data structures are unusual for lower
bounds, in that the difficult goal seems to be proving {\em lower}
lower bounds for natural problems.)

A significant break-through occured in SODA'09, when
Golynski~\cite{golynski09lb} showed a lower bound of $r\cdot t^2
= \Omega(n)$ for the problem of storing a permutation and querying
$\pi(\cdot)$ and $\pi^{-1}(\cdot)$. This quadratic trade-off is tight
for storing a permutation and its inverse. Golynski's technique is
based on the inherent difficulty of storing a function and its inverse
without doubling the space. However, due to the particular attention
it pays to inverses, it is unclear how it could generalize to problems
like rank.

In this paper, we make further progress on getting lower bounds for
natural problems, and analyze one of the central problems in succinct
data structures. It is reasonable to hope that our lower bound
technique will generalize to many other problems, given the many
applications of rank queries.

\section{The Proof}

\subsection{An Entropy Bound}

The structure of the rank problem is not particularly important in the lower bound proof. All that is needed is an inequality on the entropy of rank queries that we describe here. Essentially, the lower bound applies to any problem which satisfies a similar entropy condition.

The possible queries come from the universe $[n]$. Imagine that this universe is divided into $k$ blocks of equal size (the remainder is ignored if $k$ doesn't divide $n$). Let $Q_\Delta \subset [n]$ be the set containing the $\Delta$-th query (counting from zero) in each block. For a set $Q$ of queries, let $\ans(Q)$ be the vector of answers to the queries in $Q$. We treat $\ans(Q)$ as a random variable, depending on the random choice of the input $A[1\twodots n]$.

\begin{lemma}    \label{lem:correl}
Let $A$ is chosen uniformly at random in $\{0,1\}^n$, and let$\Delta$ and any $Q^\star \subseteq Q_\Delta$ be arbitrary. Then, for any event $\evt$ with $\Pr[\evt] = 2^{-\eps |Q^\star|}$ for a small enough constant $\eps$, we have:
\[ H(\ans(Q_0) \mid \evt) + H(\ans(Q^\star) \mid \evt) - H(\ans(Q_0), \ans(Q^\star) \mid \evt) = \Omega(|Q^\star|) \]
\end{lemma}

\begin{proof}
Let us ignore the conditioning on $\evt$ for now. The lemma says that representing the answers to the queries $Q_0$ and (a subset of) $Q_\Delta$ separately loses $\Omega(1)$ bits of entropy per block compared to the optimal joint encoding.

Let $h_m$ be entropy of the binomial distribution on $m$ unbiased trials. The entropy $H(\ans(Q_0))$ is exactly equal to $k \cdot h_{n/k}$: the answer of a query minus the answer of the previous is exactly a binomial on $n/k$ random bits. In all blocks that do not contain an element of $Q^\star$, the contribution of the block in $H(\ans(Q_0))$ is cancelled by its contribution in $H(\ans(Q_0), \ans(Q^\star))$.

Blocks that contain an element from $Q^\star$ (except the first block) contribute:
\begin{itemize}
\item $h_{n/k}$ to $H(\ans(Q_0))$;
\item at least $h_{n/k}$ to $H(\ans(Q^\star))$. The contribution is more if the previous block did not contain an element from $Q^\star$; 
\item exactly $h_\Delta + h_{n/k - \Delta}$ to $H(\ans(Q_0), \ans(Q^\star))$.
\end{itemize}

Thus, the block contributes $2 h_{n/k} - h_\Delta - h_{n/k - \Delta}$ to the sum. Using the known estimation $h_m = \frac{1}{2} \ln(\frac{\pi e}{2} m) + O(\frac{1}{m})$, this quantity is minimized when $\Delta = \frac{n}{2k}$, and is always at least $\ln 2 - o(1)$.

The fact that conditioning on $\evt$ does not change the result comes from a standard independence trick in lower bounds. We decomposed $H(\ans(Q_0)) + H(\ans(Q^\star) - H(\ans(Q_0), \ans(Q^\star))$ as the sum over $Q^\star$ independent variables (essentially%
\footnote{The careful reader has probably noticed that we actually decomposed it into \emph{two} sums, each of which has $Q^\star$ terms independent among themselves; however, the sums are dependent. We are subtracting the entropy of sub-blocks of size $\Delta$ from the entropy of blocks of size $n/k$ in the first sum; and the entropy of sub-blocks of size $n/k - \Delta$ from the entropy of blocks of size $n/k$ in the second sum. The analysis proceeds by union bound over the two sums.
}
). Each component was $\Omega(1)$ with constant probability. By a Chernoff bound, the sum is $\Omega(|Q^\star|)$ with probability $2^{-\Omega(|Q^\star|)}$. Thus, even if we condition on an event of probability $2^{\eps |Q^\star|}$, the sum must remain $\Omega(|Q^\star|)$ with overwhelming probability.
\end{proof}

\subsection{Cell-Probe Elimination}

To support the induction in our proof, we augment the cell-probe model with \emph{published bits}. These bits represent a memory of bounded size which the query algorithm can examine \emph{at no cost}. Like the regular memory (which must be examined through cell probes), the published bits are initialized at construction time, as a function of the input $A[1\twodots n]$. Observe that if we have $n$ published bits, the problem can be solved trivially.

Our proof will try to publish a small number of cells from the regular memory which are accessed frequently. Thus, the complexity of many queries will decrease by at least one. The argument is then applied iteratively: the cell-probe complexity decreases, as more and more bits are published. If we arrive at zero cell probes and less than $n$ published bits, we have a contradiction.

Let $\probes(q)$ be the set of cells probed by query $q$; this is a random variable, since the query can be adaptive. Also let $\probes(Q) = \bigcup_{q\in Q} \probes(q)$.

The main technical result in our proof is captured in the following lemma, the proof of which appears in the next section:
\begin{lemma} \label{lem:tech}
Assume a data structure uses $P = o(n)$ published bits, and at most $n$ memory bits. Break the queries into $k = \gamma\cdot P$ blocks, for a large enough constant $\gamma$. Then:
\[ \Pr_{A,q\in [n]} \big[ \probes(q) \cap \probes(Q_0) \ne \emptyset 
    \big] = \Omega(1) \]
\end{lemma}

The lemma shows that $\probes(Q_0)$ are a good set of cells to publish, since a constant fraction of the queries probe at least one cell from this set.

Completing the proof is now easy. If the data structure has redundancy $r$, begin by publishing some arbitrary $P_0 = r$ bits, to satisfy the condition that there are at most $n$ bits in regular memory.

In step $i=0,1,2\dots$, we let $k_i = \gamma \cdot P_i$, and publish the cells in $\probes(Q_0)$, together with their address. The number of published bits increases to $P_{i+1} = k_i \cdot (w + O(\lg n)) = O(P_i w)$. The cell-probe complexity of an average query decreases by $\Omega(1)$.

Since the average case complexity cannot go below zero, the number of iterations that we are able to make must be $O(t)$. The only reason we may fail to make another iteration is a violation to the lemma's condition $P=o(n)$. Thus, $P_{O(t)} = \Omega(n)$, that is $r \cdot w^{O(t)} \ge n$. This is the desired trade-off.

\subsection{An Encoding Argument}

In this section, we prove Lemma~\ref{lem:tech}. Our proof is an encoding argument: we show that, if the conclusion of the lemma failed, we could encode a uniformly random $A$ using strictly less than $n$ bits.

Let $P$ and $k$ be as in our lemma's statement, and assume $\Pr_{A,q \in [n]}[\probes(q) \cap \probes(Q_0) \ne \emptyset ] \le \eps$, for a small enough constant $\eps$. We thus know that a random query is very likely to probe cells not in $\probes(Q_0)$.

By averaging, there exists a $\Delta \in \{1, \dots, n/k \}$ such that
$ \Pr_{A, q\in Q_\Delta} [ \probes(q) \cap \probes(Q_0) \ne \emptyset
] \le \eps. $ We are only going to concentrate on the queries in $Q_\Delta$. 

More specifically, we are going to concentrate on the queries that probe no cell from $\probes(Q_0)$: $Q^\star = \{ q \in Q_\Delta \mid \probes(q) \cap \probes(Q_0) = \emptyset \}$. Note that $\E_A [|Q^\star|] \ge (1-\eps) k$.

Intuitively speaking, our contradiction is found as follows. The answers to queries $Q_0$ must be encoded in the cells $\probes(Q_0)$. The answers to queries $Q^\star$ must be encoded in the cells $\probes(Q^\star)$, which, by definition, is disjoint from $\probes(Q_0)$. But the answers $\ans(Q_0)$ and $\ans(Q^\star)$ are highly correlated (by Lemma~\ref{lem:correl}). Thus, if the two answers are written in disjoint sets of cells, a lot of entropy is being wasted, which is impossible for a succinct data structure.

\paragraph{The footprint.}
We first formalize the intuitive notion of ``the contents of cells $\probes(Q)$.'' Define the footprint $\foot(Q)$ of a query set $Q$ by the following algorithm. We assume the published bits are known in the course of the definition. Enumerate queries $q \in Q$ in increasing order. For each query, simmulate its execution one cell probe at a time. If a cell has already been included in the footprint, ignore it. Otherwise, append the contents (but not the address) of the new cell in the footprint. Observe that $\foot(Q)$ is a string of exactly $|\probes(Q)| \cdot w$ bits.

We observe that $\ans(Q)$ is a function of $\foot(Q)$ and the published bits. Indeed, we can simmulate the queries in order. At each step, we know how the query algorithm acts based on the published bits and the previously read cells. Thus, we know the address of the next cell to be read. We can check whether the cell was already in the footprint (since we also know the address of previous cells). If not, we read the next $w$ bits of the footprint, which are precisely the contents of this cell, and continue the simulation.

\paragraph{The encoding.}
Our encoding for the array $A$ will consist of the following:
\begin{enumerate*}
\item the published bits ($P$ bits). Denote these bits by the random variable $\pub$.

\item the identity of the set $Q^\star$ as a subset of $Q_\Delta$. This uses $O\big( \lg \binom{k}{|Q^\star|} \big) = O\big( \lg \binom{k}{k - |Q^\star|} \big)$ bits. By submodularity, the average length of this component is on the order of:
\[ \E\Big[ \lg \binom{k}{k-|Q^\star|} \Big] 
~\le~ \lg \binom{k}{\E[k-|Q^\star|]} 
~\le~ \lg \binom{k}{\eps k} ~=~ k \cdot O(\eps \lg \tfrac{1}{\eps})
\]

\item the answers $\ans(Q_0 \cup Q^\star)$, encoded jointly. Using Huffman coding, this requires $H(\ans(Q_0 \cup Q^\star)) + O(1)$ bits on average.

\item the footprint $\foot(Q_0)$, encoded optimally given the knowledge of $\ans(Q_0)$ and the published bits. This takes $H(\foot(Q_0) \mid \ans(Q_0), \pub) + O(1)$ bits on average.

\item the footprint $\foot(Q^\star)$, encoded optimally given the knowledge of $Q^\star, \ans(Q^\star),$ and the published bits. This takes $H(\foot(Q^\star) \mid Q^\star, \ans(Q^\star), \pub) + O(1)$ bits on average.

\item all cells outside $\probes(Q_0) \cup \probes(Q^\star)$, included verbatim with $w$ bits per cell. As noted above, the cell addresses $\probes(Q_0)$ and $\probes(Q^\star)$ can be decoded from $\foot(Q_0)$, respectively $\foot(Q^\star)$, and the published bits. Thus, we know exactly which cells to include in this component. This part takes $n - \E[|\probes(Q_0)| + |\probes(Q^\star)|] \cdot w$ bits on average.
\end{enumerate*}

Observe that this encoding includes the published bits and all cells in the memory (though the cells in $\probes(Q_0)$ and $\probes(Q^\star)$ are included in a compressed format). Thus, all $n$ queries can be simmulated. If all $n$ answers are known, the array $A$ can be decoded. Thus, this is a valid encoding of $A$.

It remains to analyze the average size of the encoding. To bound item 4., we can write:
\[ H(\foot(Q_0) \mid \ans(Q_0), \pub)
~=~ H(\foot(Q_0), \ans(Q_0), \pub) - H(\ans(Q_0), \pub) \]
But $H(\foot(Q_0), \ans(Q_0), \pub) = H(\foot(Q_0), \pub)$, since the answers can be decoded from the footprint and the published bits. Now note that $H(\foot(Q_0), \pub) \le \E[|\probes(Q_0)|]\cdot w + P$, since this is the size in bits of the footprint and the published bits. Finally, note that $H(\ans(Q_0), \pub) \ge H(\ans(Q_0))$. Thus:
\[ H(\foot(Q_0) \mid \ans(Q_0), \pub) ~\le~
\E[|\probes(Q_0)|]\cdot w + P - H(\ans(Q_0))
\]
Similarly, item 5.~is bounded by:
\[ H(\foot(Q^\star) \mid Q^\star, \ans(Q^\star), \pub)
~\le~ \E[|\probes(Q^\star)|]\cdot w + P + 
    k\cdot O(\eps \lg \tfrac{1}{\eps}) - H(Q^\star, \ans(Q^\star))
\]
Summing up all components, our encoding has expected size:
\begin{equation}  \label{eq:size}
n + 3P + k\cdot O(\eps \lg \tfrac{1}{\eps}) + H(\ans(Q_0), \ans(Q^\star)) - H(\ans(Q_0)) - H(Q^\star, \ans(Q^\star))
\end{equation}

We can now rewrite:
\begin{eqnarray*}
& & H(\ans(Q_0)) + H(Q^\star, \ans(Q^\star)) - H(\ans(Q_0), \ans(Q^\star)) \\
&\ge&
H(\ans(Q_0) \mid Q^\star) + H(\ans(Q^\star) \mid Q^\star) + H(Q^\star)
- H(\ans(Q_0), \ans(Q^\star), Q^\star) \\
&=&
H(\ans(Q_0) \mid Q^\star) + H(\ans(Q^\star) \mid Q^\star) + H(Q^\star)
- H(\ans(Q_0), \ans(Q^\star) \mid Q^\star) - H(Q^\star) \\
&=&
H(\ans(Q_0) \mid Q^\star) + H(\ans(Q^\star) \mid Q^\star)
- H(\ans(Q_0), \ans(Q^\star) \mid Q^\star) \\
&=& \E_{\fixQ} \big[ H(\ans(Q_0) \mid Q^\star = \fixQ) 
+ H(\ans(\fixQ) \mid Q^\star = \fixQ)
- H(\ans(Q_0), \ans(\fixQ) \mid Q^\star = \fixQ) \big]
\end{eqnarray*}

We can now apply Lemma~\ref{lem:correl} for any fixed $\fixQ$ and the event $\evt =\{ Q^\star = \fixQ\}$. Note that the density $\Pr[\evt]$ is $2^{-k \cdot \Omega(\eps \lg \frac{1}{\eps})}$ which constant probability over the choice of $\fixQ$. Thus, the lemma applies for small enough $\eps$. We conclude that $H(\ans(Q_0) \mid \evt) + H(\ans(\fixQ) \mid \evt) - H(\ans(Q_0), \ans(\fixQ) \mid \evt) = \Omega(k)$ with constant probability over $\fixQ$. Thus, the expectation is also $\Omega(k)$.

Plugging our result into \eqref{eq:size}, the size of the encoding becomes $n + 3P + k \cdot O(\eps \lg \tfrac{1}{\eps}) - \Omega(k)$. Setting $k = \gamma P$ for a large constant $\gamma$, and $\eps$ a small enough constant, the negative $\Omega(k)$ term is double the positive terms. Thus, the encoding size is $n - \Omega(k)$, a contradiction.
{
\bibliographystyle{alpha} 
\bibliography{general}
}

\end{document}